\documentclass[lettersize,journal]{IEEEtran}
\usepackage{amsmath,amsfonts}
\usepackage{algorithmic}
\usepackage{array}
\usepackage{textcomp}
\usepackage{stfloats}
\usepackage{url}
\usepackage{verbatim}
\usepackage{makecell}
\usepackage{graphicx}
\usepackage{xcolor}
\usepackage{amssymb}
\usepackage{booktabs}
\usepackage{subfigure}

\makeatletter

\newcommand{\Rmnum}[1]{\expandafter\@slowromancap\romannumeral #1@}
\makeatother

\usepackage[sort,compress]{cite}
\usepackage[
colorlinks=true,
linkcolor=blue,
citecolor=blue,
urlcolor=blue
]{hyperref}

\usepackage{multirow}
\def\BibTeX{{\rm B\kern-.05em{\sc i\kern-.025em b}\kern-.08em T\kern-.1667em\lower.7ex\hbox{E}\kern-.125emX}}
\usepackage{balance}
\begin{document}
\title{Sensing-Assisted LoS/NLoS Identification in Dynamic UAV Positioning Systems}
\author{Huijuan~Qiao,~\IEEEmembership{Graduate~Student~Member,~IEEE}, Lu~Bai,~\IEEEmembership{Senior~Member,~IEEE}, Mingran~Sun,~\IEEEmembership{Graduate~Student~Member,~IEEE}, Mengyuan~Lu,~\IEEEmembership{Graduate~Student~Member,~IEEE},\\ Jiajing~Chen, and Xiang~Cheng,~\IEEEmembership{Fellow,~IEEE}
 % <-this % stops a space

 %\thanks{Manuscript received 9 July 2025; revised 12 September 2025; accepted 17 December 2025. This work was supported in part by the National Natural Science Foundation of China under Grant 62371273, Grant 62125101, Grant 62341101, Grant 62571007, and Grant 62501019; in part by the Taishan Scholars Program under Grant tsqn202312307; in part by the Young Elite Scientists Sponsorship Program by CAST under Grant YESS20230372; in part by the Shandong Natural Science Foundation under Grant ZR2023YQ058 and ZR2025MS1019; in part by the New Cornerstone Science Foundation through the XPLORER PRIZE; in part by the Xiaomi Young Talents Program; in part by the open research fund of National Mobile Communications Research Laboratory, Southeast University (No. 2025D04); in part by the Fundamental Research Funds for the Central Universities, Peking University; in part by the Beijing Natural Science Foundation under Grant 4254067; in part by the China National Postdoctoral Program for Innovative Talents under Grant BX20240007; and in part by the China Postdoctoral Science Foundation under Grant 2024M760111. The authors would like to thank Shiliang Lu for the help in the dataset construction at mmWave and low-THz frequency bands under National Stadium scenario via Wireless InSite simulation platform. (\textit{Corresponding author: Xiang Cheng.})}
 
%白老师单位(投稿阶段只写山大，定稿再加单位)
\thanks{H.~Qiao and M.~Lu are with the Joint SDU-NTU Centre for Artificial Intelligence Research (C-FAIR), Shandong University, Jinan 250101, China, and also with the School of Software, Shandong University, Jinan 250101, China (e-mail: huijuanqiao@mail.sdu.edu.cn; mengyuanlu@mail.sdu.edu.cn).}
\thanks{L.~Bai is with the Joint SDU-NTU Centre for Artificial Intelligence Research (C-FAIR), Shandong University, Jinan 250101, China (e-mail: lubai@sdu.edu.cn).}
\thanks{M.~Sun and X.~Cheng are with the State Key Laboratory of Photonics and Communications, School of Electronics, Peking University, Beijing 100871, China (email: mingransun@stu.pku.edu.cn; xiangcheng@pku.edu.cn).}
\thanks{J.~Chen is with the School of Information and Intelligence Science, Donghua University, Shanghai 201620, China (email: chenjiajing@tongji.edu.cn).}
}

% The paper headers
\markboth{IEEE Transactions on Communications, vol. xx, no. xx, XX 2026}{}

\maketitle

\begin{abstract}
In this paper, a sensing-assisted non-line-of-sight (NLoS) identification method for dynamic uncrewed aerial vehicle (UAV) positioning is proposed for the first time. For urban UAV-to-ground scenarios, a new multi-modal sensing-communication integrated dataset is constructed to support line-of-sight (LoS)/NLoS identification, covering two typical urban scenarios and a wide range of flight altitudes. Based on the constructed dataset, a novel dual-input feature fusion network is proposed, which addresses the challenge of heterogeneous representations between RGB images and channel impulse response (CIR) data to enable the joint extraction and fusion of sensing and communication features for LoS/NLoS identification. Simulation results show that the identification accuracy can reach up to 97.69\%, while achieving an improvement of at least 3.59\% compared to traditional CIR-only and RGB-only methods. Moreover, strong few-shot generalization is observed, as the proposed method stabilizes and approaches full-sample performance with fewer than 200 target samples and exceeds traditional CIR-only and RGB-only methods with fewer than 100 target samples in all cross-scenario and cross-altitude experiments. Even under Gaussian noise with a variance of 0.35 applied to RGB images, the accuracy degradation remains approximately 0.5\%. By utilizing the proposed LoS/NLoS identification method, the error of trilateration positioning can be reduced by approximately 70\% in a crossroad scenario, verifying the utility of the proposed method.
\end{abstract}


\section{Introduction}
\IEEEPARstart{W}{ith} sixth-generation (6G) communication technology extends into the low-altitude domain, uncrewed aerial vehicle (UAV)-to-ground communications have become an essential component of low-altitude transportation \cite{low-altitude}. Thanks to high mobility and flexibility, as well as the improved communication quality provided by predominantly line-of-sight (LoS) air-to-ground links, UAVs can be deployed to provide various types of services in dense urban scenarios \cite{UAVlocalization}. For UAVs in low-altitude transportation systems, positioning technology serves as a fundamental basis for ensuring safety and efficient operation \cite{UAVpositioning}. For example, in UAV-assisted emergency rescue missions, accurate positioning of the UAV ensures its ability to effectively penetrate complex environments such as smoke or ruins and accurately reach the locations of trapped individuals \cite{Rescue}. In addition to supporting safe operation, dynamic UAV positioning can further improve network coverage \cite{networkcoverage}. In positioning technologies, the trilateration algorithm is a widely employed geometric method and serves as the foundation for numerous contemporary positioning systems. When time of arrival (ToA) and similar measurements are available, trilateration becomes a suitable and effective positioning method \cite{positioning,trilateration}. Nevertheless, in UAV trilateration positioning, non-line-of-sight (NLoS) paths can introduce positive bias in distance estimation, significantly affecting overall positioning accuracy \cite{urban}. In urban scenarios characterized by dense high-rise buildings and complex traffic structures, there are numerous NLoS propagation paths which pose significant challenges to trilateration-based positioning. Therefore, the ability to efficiently and accurately identify LoS and NLoS conditions in complex dynamic urban scenarios is critical to enhancing the performance of positioning systems.

LoS/NLoS identification has attracted widespread attention from researchers. For the conventional LoS/NLoS identification methods, the primary purpose is to assist in positioning or channel modeling in specific scenarios. Specifically, the conventional LoS/NLoS identification methods can be categorized into three main categories, i.e., the manual observation-based method, the deterministic threshold-based method, and the statistical analysis-based method \cite{categories}. For the manual observation-based method, LoS/NLoS identification is performed through human visual observation. In \cite{traditional1}, the LoS and NLoS regions were identified from the recorded measurement route and map of each scenario. However, the identification method is time-consuming and exhausting for observers. To address this problem, the deterministic threshold-based method utilizes deterministic thresholds based on channel characteristic parameters to identify LoS and NLoS. Specifically, the authors in \cite{traditional2} employed the deviation between the received signal strength indication (RSSI) measurement and its predicted value as the discriminant criterion. Although the method mentioned in \cite{traditional2} is computationally simple and fast, it heavily relies on the precise setting of the threshold and environmental stability. To address this problem, the statistical analysis-based method was proposed. In \cite{traditional3}, the authors realized LoS/NLoS identification based on a binary hypothesis test by characterizing the statistics of the channel impulse response (CIR), such as kurtosis and coherence bandwidth. The method mentioned in \cite{traditional3} is relatively less sensitive to environmental changes. However, the identification performance of the method is closely related to the selected statistical characteristics. Moreover, it is limited to two-dimensional (2D) single-layer indoor scenarios. In summary, the aforementioned methods exhibit good LoS/NLoS identification performance but are primarily designed for specific scenarios. In addition, insufficient robustness in dynamic scenarios and poor generalization to new scenarios make them difficult to meet the demands of UAV positioning. Therefore, the aforementioned methods are not well suited to high dynamic UAV positioning in complex urban scenarios.

Fortunately, artificial intelligence (AI) has achieved rapid development and shown good performance in solving identification problems. Therefore, it has been applied to LoS/NLoS identification, such as support vector machines (SVM), random forests (RF), artificial neural networks (ANN), etc. The authors in \cite{SVM1} and \cite{SVM2} utilized SVM-based machine learning methods for LoS/NLoS identification. In \cite{SVM1,SVM2}, CIR and various statistical features were employed for LoS/NLoS identification. Machine learning classifiers, which includes linear discriminant analysis (LDA) and SVM were utilized, and the impact of feature combination on identification accuracy was analyzed. However, the works in \cite{SVM1, SVM2} are aimed at indoor scenarios, which differ from the complex urban scenarios. For dynamic vehicle-to-vehicle (V2V) scenarios in campus and surrounding environments, the authors in \cite{threeM} utilized three machine learning methods, i.e., SVM, RF, and ANN, based on several static and time-varying features of CIR for LoS/NLoS identification. Authors in \cite{CNN} proposed a NLoS identification algorithm by utilizing a convolutional neural network (CNN) based on the power-angle-spectrum (PAS) in a dynamic suburban crossroad V2V scenario. The methods in \cite{threeM, CNN} both utilized PAS and demonstrated effectiveness in dynamic scenarios. However, due to their reliance on PAS, the antenna array may impact adaptability to three-dimensional (3D) scenarios, and their implementation is limited to identifying the channel state for a single receiving point. In summary, the aforementioned AI-based LoS/NLoS identification methods in \cite{SVM1, SVM2, threeM, CNN} are limited to indoor scenarios or are easily affected by feature selection, and identify the channel state for only a single receiving point. Furthermore, the aforementioned methods are limited to utilizing radio frequency information to identify LoS/NLoS, which ignore the sensing information, e.g., RGB images, contained in the physical environment. The geometric structure information of the physical environment embedded in RGB images can directly assist in identifying the LoS/NLoS status of propagation paths.

To overcome these limitations, a new concept Synesthesia of Machines (SoM) was proposed in \cite{SoM}. SoM refers to intelligent multi-modal sensing-communication integration. Under the SoM framework, multi-modal intelligent channel modeling (MMICM) was introduced to intelligently fuse multi-modal information for accurate channel characterization \cite{MMICM, mmicm}. MMICM inspired us to explore the correlation between the physical environment and electromagnetic propagation characteristics for LoS/NLoS identification. As a result, a sensing-assisted LoS/NLoS identification method based on SoM for dynamic UAV positioning is proposed for the first time. The main contributions and novelties of this paper are as follows.
\begin{enumerate}
    \item A novel sensing-assisted LoS/NLoS identification method based on SoM for dynamic UAV positioning is proposed for the first time, which supports accurate and real-time UAV positioning for low-altitude transportation. Due to the proposed method captures the correlation between the physical environment and electromagnetic propagation characteristics, it demonstrates strong generalization across diverse scenarios and flight altitudes, thereby offering a robust and efficient solution for enhancing positioning.
    \item To capture the correlation between the physical environment and electromagnetic propagation characteristics, a new multi-modal sensing-communication integrated dataset in UAV-to-ground scenarios is constructed. The constructed dataset contains a total of 7,840 RGB images, 7,840 CIR matrices, and 7,840 LoS/NLoS identification label matrices, covering two urban scenarios under 28 GHz, i.e., crossroad and wide lane scenarios, and three flight altitudes, including 50 m, 63.3 m, and 70 m.
    \item To achieve the fusion of sensing image and communication data, a dual-input feature fusion identification network integrating sensing and communication is developed to LoS/NLoS identification for the first time. The novel feature extraction module based on Vision Transformer (ViT) and CNN is designed to extract feature maps, which achieves spatial alignment of the dual modal features. Building on the achieved alignment, a newly designed fusion feature extraction and classification module leverages convolution to deeply integrate the dual modal features and output a classification probability matrix, ultimately achieving accurate and robust LoS/NLoS identification.
    \item Simulation results demonstrate that the proposed method, which fuses RGB image and CIR, achieves excellent accuracy in full-sample trained LoS/NLoS identification. Across different UAV flight altitudes and scenarios, the proposed method achieves identification accuracies of no less than 95.49\%. Improvements of approximately 4\% to 15\% are achieved compared to traditional CIR-only and RGB-only methods. Furthermore, in cross-scenario and cross-altitude experiments, the method stabilizes and approaches full-sample performance with fewer than 200 target samples and exceeds traditional CIR-only and RGB-only methods with fewer than 100 target samples. Additionally, when Gaussian noise is gradually added to the RGB images, the accuracy decreased by only approximately 0.5\% at a noise level of 0.35, the fused method demonstrates strong robustness against such disturbances. The proposed method can also assist in reducing the positioning error by approximately 70\% compared to random method.
\end{enumerate}

The remainder of the paper is organized as follows. Section \Rmnum{2} presents the construction of the new multi-modal sensing-communication integrated dataset. In Section \Rmnum{3}, the scheme of the proposed LoS/NLoS identification method is presented. Subsequently, the accuracy, generalization, and performance in positioning assistance of the proposed method are evaluated in Section \Rmnum{4}. Finally, conclusions are presented in Section \Rmnum{5}.

\section{Dataset Construction}

\begin{figure*}[h]
\centering
\includegraphics[scale = 0.53]{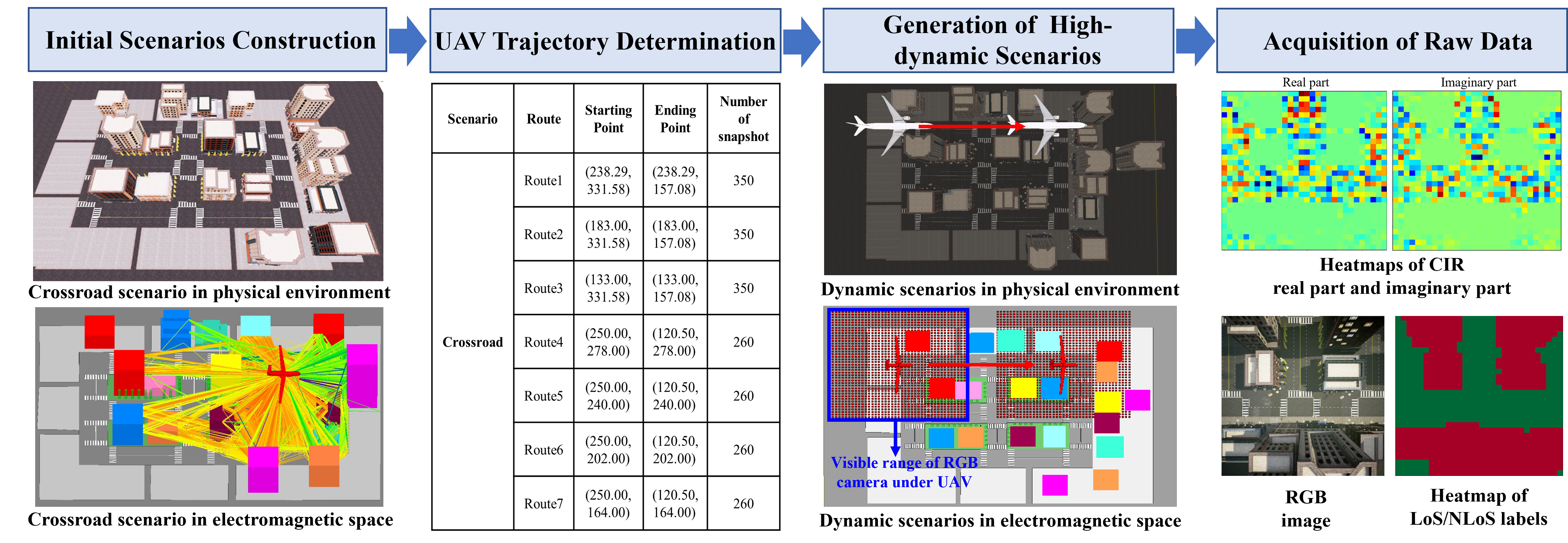}
\caption{Processing flow of dataset construction for the crossroad scenario.}
\label{dataset}
\end{figure*}

To support the research on the identification of LoS/NLoS in UAV-to-ground scenarios, a new dataset that integrates the sensing information and communication information of UAV-to-ground communications is constructed. The dataset contains RGB images, CIR matrices, and propagation path classification labels indicating LoS and NLoS collected in two urban scenarios, i.e., crossroad and wide lane scenarios, and three UAV flight altitudes, i.e., 50 m, 63.3 m, and 70 m. The inclusion of CIR matrices allows the dataset to support more extensive channel modeling and sensing research beyond LoS/NLoS identification. Specifically, the new multi-modal sensing-communication integrated dataset contains a total of 7,840 RGB images, 7,840 CIR matrices, and 7,840 LoS/NLoS identification label matrices. Based on the in-depth integration and precise alignment of physical environment and electromagnetic space across AirSim \cite{AirSim} and Wireless InSite \cite{WI}, the new sensing-communication integrated dataset is constructed. As shown in Fig. \ref{dataset}, the construction of the dataset involves four key steps, i.e., initial scenarios construction, UAV trajectory determination, generation of high-dynamic scenarios, and acquisition of raw data. The detailed process of dataset construction is described below.

\subsection{Initial Scenario Construction}

For the crossroad scenario, the first step of dataset construction is to construct the initial scenario. As shown in Fig. \ref{dataset}, the urban crossroad scenario contains 19 buildings of different heights and 6 streets, with 3 streets in both horizontal and vertical directions. The area of the whole scenario is $200 \times 260 \, \text{m}^2$. In the crossroad scenario, a UAV equipped with an RGB camera and a transmitter (Tx) antenna is placed above the urban scene, while a grid of receiver (Rx) antennas is laid out within the visible ground range of the RGB camera as shown in Fig. \ref{dataset}.

\begin{figure*}[h]
\centering
\includegraphics[scale = 0.55]{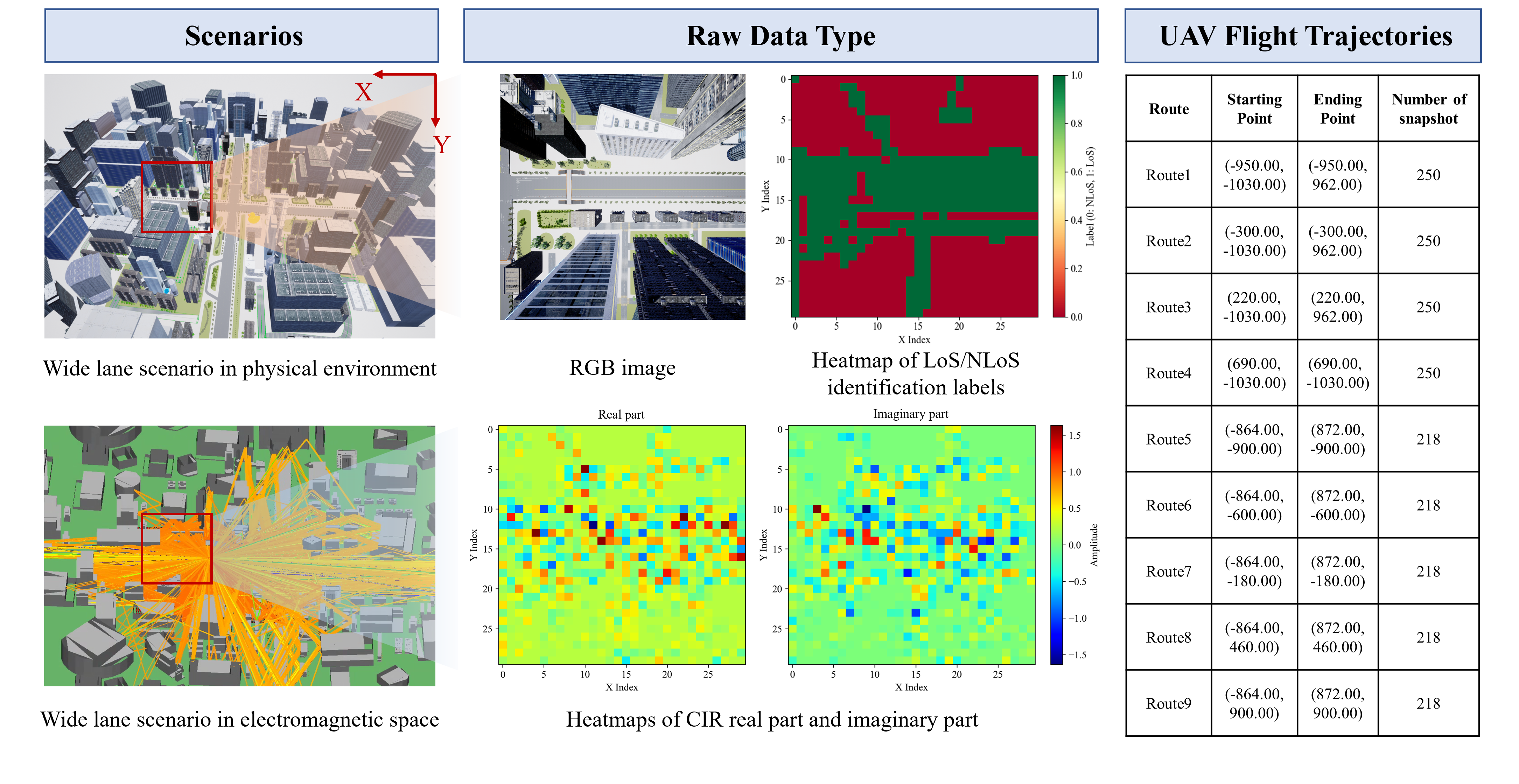}
\caption{The wide lane scenario of the constructed dataset.}
\label{scenario}
\end{figure*}

During dataset construction, it is essential to achieve in-depth integration of sensing and communication, as well as precise alignment of the physical environment and electromagnetic space. The same buildings, vegetation, and UAV are utilized for both sensing and communication data acquisition to ensure consistency between the physical environment and electromagnetic space. For sensing data acquisition, an RGB camera is equipped at the bottom of the UAV to capture RGB images. Meanwhile, a single antenna is mounted on the UAV as the Tx, and a $30 \times 30$ grid of Rxs is arranged on the ground for communication data collection. Initial scenarios at different flight altitudes can be obtained by adjusting the altitude of the UAV and its onboard equipment. In addition to the crossroad scenarios, wide lane scenarios are also constructed. Fig. \ref{scenario} illustrates a wide lane scenario, where the buildings are more densely distributed compared to the crossroad scenarios.

\subsection{Generation of High-Dynamic Scenarios Based on Determined UAV Trajectories}

To cover a wide range of UAV operational behaviors and ensure data diversity, flight trajectories are predefined and scenarios are generated in batches. Based on the initial crossroad scenario, multiple UAV trajectories over urban roads and buildings for the UAV are set, which ensures trajectory diversity and system robustness. As shown in Fig. \ref{dataset}, the UAV in crossroad scenarios is set to fly along seven predefined routes over the city at a constant speed, while the Rx grid on the ground is positioned below the UAV. To enable data acquisition along the predefined UAV trajectories in both physical environment and electromagnetic space, the UAV position is moved snapshot by snapshot to generate high-dynamic scenarios in batches, thereby collecting RGB images and CIR matrices at different time instances. During the scenario generation process, the UAV maintains the same coordinates in each snapshot for both sensing and communication data collection. The trajectories in the crossroad scenario include 2090 snapshots of different positions of the UAV during the movement process, with 350 snapshots for each of routes 1-3 and 260 snapshots for each of routes 4-7. Trajectory diversity ensures data richness, which supports robust analysis under different flight behaviors and spatial layouts. Following the same procedure as in the crossroad scenario, the UAV in the wide lane scenario follows 9 predefined trajectories listed in Fig. \ref{scenario}.

\subsection{Acquisition of Raw Sensing and Communication Data}

After the batch setup of high-dynamic scenarios, the next step is to collect multi-modal sensing and communication data. Taking the crossroad scenario with a UAV flight altitude of 63.3 m as an example, a total of 2090 RGB images with a resolution of $1080 \times 1080$ are obtained by the RGB camera equipped at the bottom of the UAV in AirSim. Meanwhile, in Wireless InSite, CIR and propagation path information are collected for 2090 Rx grids, covering a total of 1,881,000 different communication paths by the Tx equipped at the bottom of the UAV. The propagation path information records the reflections, scatterings, and other interactions occurring when signals encounter obstacles during transmission between the Tx and the Rx. Based on the propagation path information, LoS/NLoS identification labels are generated for the propagation paths between the UAV and each Rx position in the Rx grid. Specifically, LoS paths are labeled as ``1", while NLoS paths are labeled as ``0". Ultimately, the constructed crossroad scenario dataset at a 63.3 m altitude includes 2090 RGB images, CIR matrices, and corresponding LoS/NLoS classification label matrices for 2090 Rx grids. For clarity, Fig. \ref{dataset} and Fig. \ref{scenario} illustrate the different types of data contained in the crossroad scenario with the UAV flying at an altitude of 63.3 m and the wide lane scenario, respectively.

For the crossroad scenarios, following the aforementioned method, UAV flights are conducted at three altitudes, i.e., 50~m, 63.3 m, and 70 m, along similar trajectories. In addition, data in the wide lane scenario are collected at a single flight altitude. As shown in Table \ref{finaldata}, the final dataset contains a total of 7,840 RGB images, 7,840 CIR matrices, and 7,840 LoS/NLoS identification label matrices, which covers three flight altitudes and two scenarios. The diversity in flight altitudes and scenarios provides robust support for subsequent performance evaluation.

%三线表
\begin{table*}[!t]
\centering
\begin{normalsize}
\caption{Key Simulation Data under Different Scenario Conditions}
\label{finaldata}
\renewcommand\arraystretch{1.5} 
\setlength{\tabcolsep}{4pt} 
\begin{tabular}{ccccc}
\toprule
\textbf{Frequency Band} & \textbf{Scenario} & \textbf{UAV Flight Altitude} & \textbf{Number of RGB Images} & \textbf{Number of CIR/Label Matrices} \\
\midrule
\multirow{4}{*}{28 GHz} & \multirow{3}{*}{Crossroad} & 50 m   & 1830 & 1830 \\
                         &                           & 63.3 m & 2090 & 2090 \\
                         &                           & 70 m   & 1830 & 1830 \\
\cmidrule(lr){2-5}   % 在第2-5列画一条短线，表示场景切换
                         & Wide Lane                 & 200 m  & 2090 & 2090 \\
\midrule
\multicolumn{3}{c}{\textbf{Total}}  & \textbf{7840} & \textbf{7840} \\
\bottomrule
\end{tabular}
\end{normalsize}
\end{table*}

\section{LoS/NLoS Identification Based on SoM}

To enable accurate and reliable LoS/NLoS identification for UAV positioning in complex urban scenarios, a novel sensing-assisted LoS/NLoS identification method is proposed. The proposed method utilizes a dual-input feature fusion classification network to explicitly model the cross‑domain correlation between the physical sensing geometry and the communication electromagnetic response. Raw multi-modal data are transformed into a unified feature space based on SoM, enabling robust LoS/NLoS identification. The network consists of a Feature Extraction Module, which is divided into two branches, i.e., ViT branch for RGB images and CNN branch for CIR matrices, and a Fusion Feature Extraction and Classification Module.

\subsection{Network Scheme}

In UAV-to-ground communication scenarios, a key challenge for LoS/NLoS identification is the limited information provided by single-modal data. To jointly leverage the physical environment information contained in sensing data and the electromagnetic space information embedded in communication data, a multi-modal fusion architecture is adopted. A major difficulty in multi-modal fusion arises from the heterogeneity between modalities, which often results in significant dimensional mismatches. To address this issue, inspired by early fusion strategies in multi-modal learning, dedicated feature extraction modules are designed for different data modalities to integrate multi-modal features and subsequently model the mapping relationship between fusion features and communication path statuses. Following the design philosophy of modality-specific feature extraction followed by collaborative fusion and reasoning, a dual-input feature fusion and classification network framework is constructed.

\begin{figure*}[h]
\centering
\includegraphics[scale = 0.50]{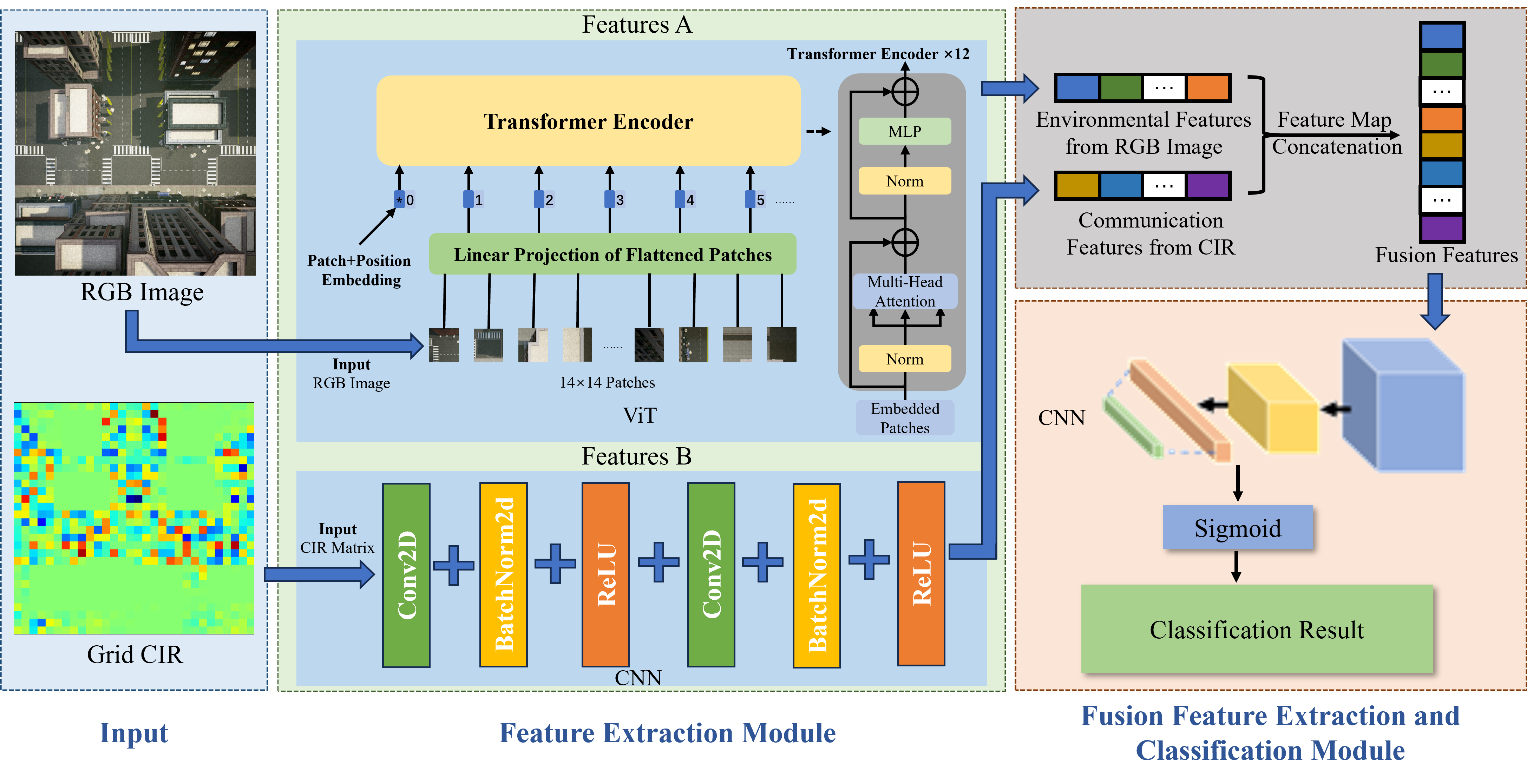}
\caption{Scheme of the proposed dual-input feature fusion network for LoS/NLoS identification.}
\label{Scheme}
\end{figure*}

As shown in Fig. \ref{Scheme}, the overall architecture comprises a dedicated feature extraction module and a fusion feature extraction and classification module. During training, labeled RGB images and CIR matrices are utilized as inputs to the model. RGB images are first resized and then processed through the ViT branch to extract environmental features. CIR matrices are processed by a CNN branch to extract communication features. The feature maps extracted from the two branches are concatenated and then passed to the fusion feature extraction and classification module to perform LoS/NLoS identification. Given the RGB image and CIR as inputs to the trained model, the predicted label matrix is obtained. The mathematical formulation of the complete network can be expressed by
\begin{equation}
\hat{y}=F_\text{fusion}(\Phi_\text{ViT}(x_\text{rgb}),\Phi_\text{CNN}(x_\text{cir}))
\end{equation}
where $x_\text{rgb} \in \mathbb{R}^{3 \times H \times W}$ represents the input RGB image, $x_\text{cir} \in \mathbb{R}^{2 \times 30 \times 30}$ denotes the preprocessed CIR matrix, $\Phi_{\text{ViT}}(\cdot)$ is the ViT-based feature extractor, $\Phi_{\text{CNN}}(\cdot)$ is the CNN-based feature extractor, $F_{\text{fusion}}(\cdot)$ represents the fusion feature extraction and classification module, and $\hat{y} \in \mathbb{R}^{30 \times 30}$ is the predicted LoS/NLoS probability map.

Multi-modal fusion can provide complementary information, which is unavailable to a single modality. By jointly leveraging environmental features extracted from RGB images and communication features obtained from CIR measurements, the proposed method learns more informative representations than single-modality approaches, which improve the identification performance.

\subsection{Feature Extraction Module}

Since the CIR at each Rx grid is a two-channel matrix with a size of $30 \times 30$, whereas an RGB image is a three-channel image with a size of $1080 \times 1080$, the data volume of the RGB modality is significantly larger than that of the CIR. Therefore, direct concatenation of the two raw modalities at the input layer would introduce substantial data redundancy \cite{Combination}. To address this issue, a feature extraction module is incorporated into the network to generate feature maps prior to fusion, which is a common design strategy in multi-modal learning. Specifically, the module consists of Features A for RGB image features extraction and Features B for CIR features extraction. Although the two modalities, i.e., RGB images and CIR matrices, differ significantly, the output of each branch is encoded into a spatial feature map of size $30 \times 30$, which facilitates effective multi-modal fusion in the subsequent stage.

\textit{1) ViT Branch (Features A):}
As shown in Fig. \ref{Scheme}, the RGB image feature extraction module utilizes a ViT \cite{ViT} , which captures long-range dependencies among all pixels through the self-attention mechanism. ViT breaks the limitations of local receptive fields and effectively captures distant occlusions, which is beneficial for understanding complex urban scenario structures. ViT splits the RGB image into patches and provides the sequence of linear embeddings of these patches as input to a Transformer, with the fewest possible modifications. The ViT model typically comprises two components, i.e., a transformer encoder that processes images to generate high-level feature representations and a classification head that maps these features to class probabilities. The classification head is replaced with an identity layer, transforming the model into a feature extraction network. This modification suppresses category probability output and instead yields raw feature vectors, enabling ViT to exclusively extract RGB image features. Finally, ViT branch outputs an RGB image feature map.

The detail architecture of the ViT branch is illustrated in Fig. \ref{Scheme}. To handle 2D images, ViT reshapes the image $x \in \mathbb{R}^{C\times H\times W}$ into a sequence of flattened 2D patches $x_p \in \mathbb{R}^{N\times (P^2\cdot C)}$, where $(H,W)$ is the resolution of the original image, $C$ is the number of channels, $(P,P)$ is the resolution of each image patch, and $N=HW/P^2$ is the resulting number of patches, which also serves as the effective input sequence length for the Transformer. The Transformer encoder maintains a constant latent vector dimension $D$ across all layers. Therefore, ViT flattens the image patches and projects them into a $D$-dimensional embedding utilizing a trainable linear projection.

The input RGB image can be expressed by $x_\text{rgb} \in \mathbb{R}^{3 \times 1080 \times 1080}$. A lightweight convolution first enhances the local texture and geometric cues
\begin{equation}
F_0 = g_\text{cnn}(x_\text{rgb})
\end{equation}
where $g_\text{cnn}(\cdot)$ consists of two convolutional layers. The feature map is resized to the ViT-required resolution by
\begin{equation}
F_\text{res} = AdaptiveMaxPool_{224 \times 224}(F_0).
\end{equation}

The ViT branch begins by splitting the image into non-overlapping $16 \times 16$ size patches
\begin{equation}
p_i = Flatten(F_\text{res}[i]) \in \mathbb{R}^{16 \times 16 \times 3}.
\end{equation}

Each patch is linearly projected into an embedding vector of dimension $D=768$

\begin{equation}
z_i^{(0)}=Ep_i+e_\text{pos,i}
\end{equation}
where $E$ is the projection matrix and $e_\text{pos,i}$ denotes positional encodings that preserve the spatial structure of the scene.

The sequence of token embeddings is processed by a stack of $L$ transformer encoder layers. At the $\ell$-th layer, the representation is updated as
\begin{equation}
Z^{(\ell)}=MSA(Z^{(\ell-1)})+FFN(Z^{(\ell-1)}),\ell=1,...,L
\end{equation}
where $MSA(\cdot)$ and $FFN(\cdot)$ correspond to multi-head self-attention and feed-forward networks.

After the final layer, the global representation is extracted by
\begin{equation}
v_\text{vit}=ViT(F_\text{res}) \in \mathbb{R}^{768}
\end{equation}
which is finally transformed through a projection layer to obtain the feature map $\mathbf{f}_\text{ViT} \in \mathbb{R}^{128 \times 30 \times 30}$.

\textit{2) CNN Branch (Features B):}

The CIR feature extraction module adopts a CNN, as shown in Fig. \ref{Scheme}. CNNs are multi-layer neural networks characterized by local connectivity and parameter sharing, and are widely utilized for feature extraction, inverse problem solving, and image classification \cite{CNN1}, \cite{CNN2}. Considering the structured spatial correlations in both the real and imaginary components of CIR matrices, the CNN architecture is employed to capture the characteristics of multi-path propagation patterns.

The input CIR matrices is expressed as 
\begin{equation}
x_\text{cir} \in \mathbb{R}^{2\times 30 \times 30}
\end{equation}
where the two channels correspond to the real and imaginary parts of the CIR. Premature spatial compression would lead to the loss of fine-grained multi-path information that is essential for identifying LoS and NLoS paths. Therefore, downsampling operations, i.e., pooling and strided convolution, are avoided in the feature extractor, which consists exclusively of convolutional layers of the same-size to preserve the complete spatial geometry of the CIR.

The CNN module contains two convolutional layers. The feature representation after the CNN is given by
\begin{equation}
\mathbf{f}_\text{CNN}=\Phi_{\text{CNN}}(x_\text{cir})
\end{equation}
where $\Phi_{\text{CNN}}$ denotes the CNN-based feature extractor. The final output is a feature map of size
\begin{equation}
\mathbf{f}_\text{CNN} \in \mathbb{R}^{128 \times 30 \times 30}
\end{equation}
which maintains full spatial resolution and serves as the refined CIR embedding for the subsequent multi-modal fusion stage.

\subsection{Fusion Feature Extraction and Classification Module}

The feature maps extracted from the Feature Extraction Module are concatenated along the channel dimension by
\begin{equation}
\mathbf{f}_{\text{fusion}} = \text{concat}(\mathbf{f}_{\text{ViT}}, \mathbf{f}_{\text{CNN}})    
\end{equation}
where $\mathbf{f}_{\text{fusion}} \in \mathbb{R}^{256 \times 30 \times 30}$ resulted in a fused feature map. This fused feature map is then fed into a deep convolutional fusion module, which consists of multiple stacked convolutional layers for further extraction of high-level features from the fused data.

Finally, a convolutional classifier processes these high-level features. This classifier gradually reduces the number of channels through a series of convolutional layers, eventually utilizing a $1 \times 1$ convolution to reduce the channels to 1. A Sigmoid activation function is applied to output a probability map of the same size as the input label, accomplishing pixel-level classification predictions. The specific equation is
\begin{equation}
\hat{y} = \text{Sigmoid}(W \mathbf{f}_{\text{fusion}}+b)
\end{equation}
where $W$ is the weight matrix to be learned, $\mathbf{f}_{\text{fusion}}$ is the input feature, $b$ is the bias term, and $\hat{y}$ is the model prediction.

The binary cross-entropy loss function \cite{BCE} is utilized to measure the training loss and update the model weights through the backpropagation algorithm. The loss function expression is
\begin{equation}
f_\text{loss} = - \sum_{i=1}^{M} \left[ y_i \log(\hat{y}_i) + (1 - y_i) \log(1 - \hat{y}_i) \right]
\end{equation}
where $M$ is the batch size, and $y_i$ is the actual grid data label.

The final decision rule of the model can be expressed as
\begin{equation}
\text{Output} = 
\begin{cases} 
1, & \text{if } \hat{y} \geq 0.5 \\ 
0, & \text{if } \hat{y} < 0.5 
\end{cases}.
\end{equation}

In summary, the fusion feature extraction and classification module for the first time integrates complementary information from both the physical environment and electromagnetic space and progressively transforms it into a unified high-level representation suitable to identify LoS/NLoS. By combining global cues captured by the ViT branch with multi-path propagation signatures extracted by the CNN branch, the proposed multi-modal fusion model achieves a more comprehensive understanding of the UAV-to-ground communication environment. This joint representation preserves both spatial consistency and modality-specific characteristics, thereby enabling the network to produce accurate LoS/NLoS identifications.

\section{Simulations and Analysis}

This section presents the identification accuracies under full-sample training. In addition the cross-scenario and cross-altitude generalization performances are evaluated. These simulations cover two urban scenarios under 28 GHz, i.e., crossroad and wide lane scenarios, and three flight altitudes, including 50 m, 63.3 m, and 70 m. Furthermore, the noise perturbation generalization and the assistance for the downstream positioning task is analyzed.

\subsection{Identification Accuracy under Full-Sample Training}

To achieve a comprehensive and systematic evaluation of the proposed identification method, simulations are conducted across two typical urban scenarios and multiple UAV flight altitudes in this subsection. Specifically, to examine the robustness of the method under altitude variations and to analyze the impact of flight altitude on LoS/NLoS identification performance, different UAV altitudes are considered in the crossroad scenario. In order to assess the robustness of the proposed method under varying structure layouts, the simulations are further extended to the wide lane scenario, where the scene structure differs significantly from that of the crossroad environment. Additionally, under different scenarios and altitudes, the proposed method is compared with identification approaches that rely solely on single-modality data, thereby highlighting the benefits of multi-modal information fusion.

\begin{table}[!t]
\centering
\begin{small}
\caption{Identification Performance of Single CIR, Single RGB Images, and Fusion Experiments}
\label{full-sample}
\renewcommand\arraystretch{1.5}
\setlength{\tabcolsep}{4pt}
\begin{tabular}{ccccc}
\toprule
\textbf{Data} & \textbf{Method} & \textbf{Scenario} & \textbf{Altitude} & \textbf{Accuracy} \\
\midrule
\multirow{4}{*}{RGB Images} & \multirow{4}{*}{CNN} & \multirow{3}{*}{Crossroad} & 50 m   & 84.06\% \\
                           &                     &                           & 63.3 m & 82.16\% \\
                           &                     &                           & 70 m   & 79.53\% \\
\cmidrule(lr){3-5}                                     % 场景切换横线
                           &                     & Wide Lane                 & 200 m  & 92.50\% \\
\midrule                                                % 数据块间横线
\multirow{4}{*}{CIR}      & \multirow{4}{*}{MLP} & \multirow{3}{*}{Crossroad} & 50 m   & 92.50\% \\
                           &                     &                           & 63.3 m & 90.12\% \\
                           &                     &                           & 70 m   & 87.32\% \\
\cmidrule(lr){3-5}
                           &                     & Wide Lane                 & 200 m  & 93.11\% \\
\midrule
\multirow{4}{*}{\makecell{RGB Images \\ + CIR}}
                           & \multirow{4}{*}{\makecell{Proposed \\ Method}}
                           & \multirow{3}{*}{Crossroad} & 50 m   & \textbf{96.09\%} \\
                           &                     &                           & 63.3 m & \textbf{96.22\%} \\
                           &                     &                           & 70 m   & \textbf{95.49\%} \\
\cmidrule(lr){3-5}
                           &                     & Wide Lane                 & 200 m  & \textbf{97.69\%} \\
\bottomrule
\end{tabular}
\end{small}
\end{table}

For the simulations, a route-based partitioning strategy is utilized to split the dataset. Specifically, in crossroad scenarios, the data from Route 2, which is illustrated in Fig. \ref{dataset}, is divided into the test set, while the data from the remaining routes is utilized for training. In the wide lane scenario, the data from Route 9, as shown in Fig. \ref{scenario}, is selected as the test set. Table \ref{full-sample} presents the identification accuracies achieved through full-sample training. The performance of the proposed method, which integrates CIR and RGB images, is compared with the RGB-only method and the CIR-only method. In the crossroad-50 m scenario, the proposed method achieves an identification accuracy of 96.09\%, which is 12.03\% and 3.59\% higher than that achieved with only RGB images or only CIR, respectively. In the crossroad-63.3 m scenario, the accuracy of the proposed method reaches 96.22\%, improving by 14.06\% and 6.1\% over the RGB-only and CIR-only methods. For the crossroad-70 m scenario, the proposed method achieves an identification accuracy of 95.49\%, outperforming the RGB-only and CIR-only methods by 15.96\% and 8.17\%, respectively. In the wide lane scenario, the proposed method achieves 97.69\% identification accuracy, with gains of 5.19\% and 4.58\% over the RGB-only and CIR-only methods. These results demonstrate that the proposed method achieves higher identification accuracy compared to single-modality methods. This performance improvement arises from the ability to effectively extract and fuse complementary features from the two modalities of the proposed method, thereby providing richer and more discriminative information for LoS/NLoS identification.

\begin{figure}[!t]
	\centering
	\subfigure[]{\includegraphics[width=0.18\textwidth]{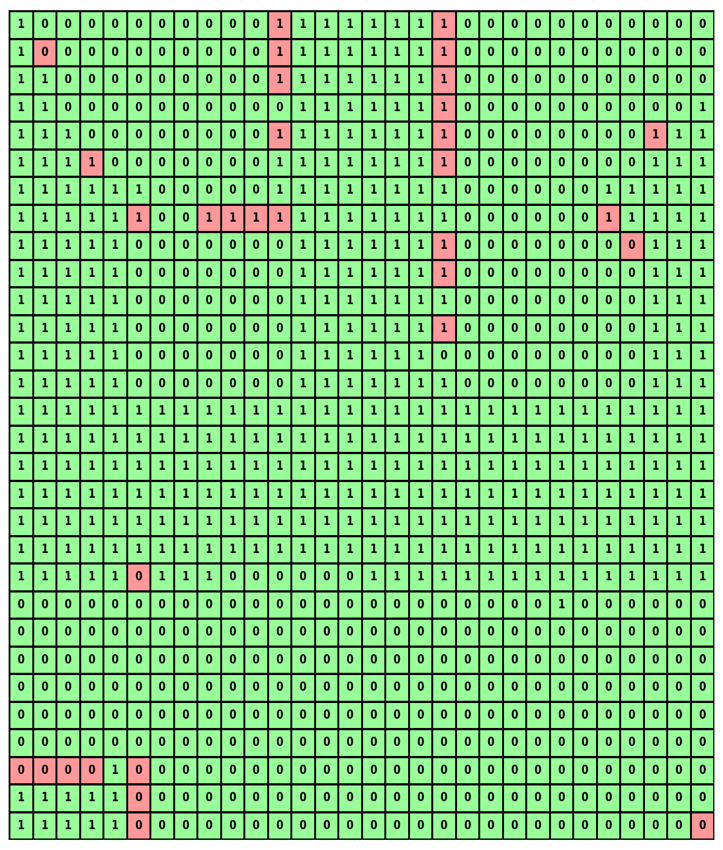}}
	\subfigure[]{\includegraphics[width=0.29\textwidth]{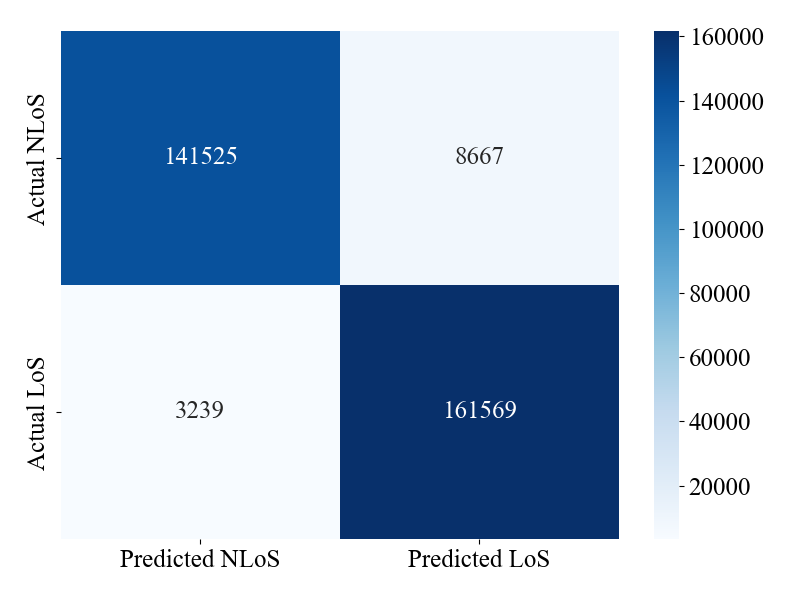}} 
	\caption{Identification results in crossroad scenario. (a) Result of a single snapshot. (b) Confusion matrix of overall test set.}
 \label{visible}
\end{figure}

Fig. \ref{visible} illustrates the LoS/NLoS identification results of the UAV during its flight in the crossroad scenario with an altitude of 63.3 m. In Fig. \ref{visible}(a), the label ``1" denotes LoS and the label ``0" denotes NLoS. Correctly identified propagation paths are highlighted in green, while incorrectly identified paths are marked in red. In the snapshot, the number of correctly identified paths is 868, which significantly outnumbers the 32 incorrectly identified ones, demonstrating that the proposed model achieves high identification accuracy. Fig. \ref{visible}(b) presents the confusion matrix of the overall test set of the crossroad-63.3 m scenario, which further evaluates the identification performance for both LoS and NLoS paths.

\begin{figure}[h]
\centering
\includegraphics[scale = 0.32]{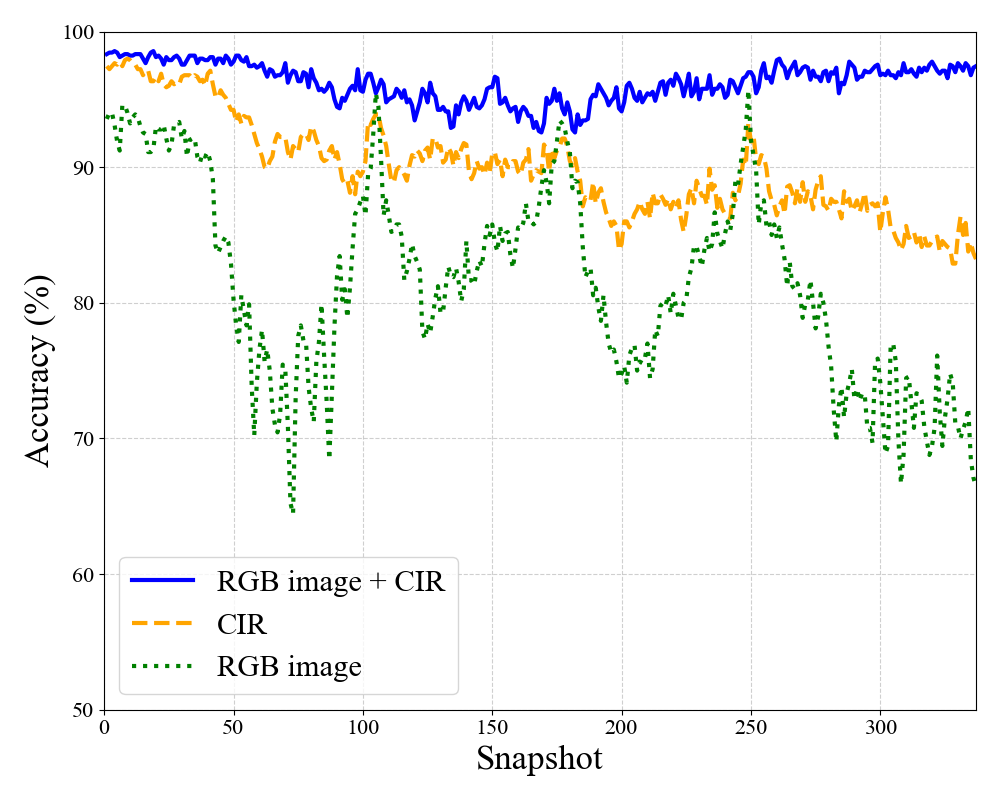}
\caption{LoS/NLoS identification accuracy of each snapshot on Route 2 in crossroad scenario.}
\label{stability}
\end{figure}

Fig. \ref{stability} shows the snapshot-level identification accuracy for the RGB-only method, the CIR-only method, and the proposed method in the crossroad-63.3 m scenario. The horizontal axis represents the UAV at different snapshots along Route 2, while the vertical axis indicates the identification accuracy at each snapshot. As can be seen from Fig. \ref{stability}, the proposed method (blue line) achieves higher and more stable identification accuracy than the other two methods. The result can be attributed to the limited environmental information contained in RGB images and CIR matrices individually, which prevents them from accurately capturing environmental variations. Consequently, the two single-modality methods exhibit accuracy peaks when the test environment resembles the training environment and accuracy valleys when the test environment deviates markedly from the training one. By exploiting the complementarity between RGB images and CIR matrices, the proposed fusion method provides a more comprehensive representation of the propagation environment and leads to improved robustness, reduced accuracy fluctuations, and superior performance over single-modality methods.

\subsection{Cross-Scenario and Cross-Altitude Generalization Performance Evaluation}

Accurate positioning is essential for UAV task execution, and LoS/NLoS identification plays a key role in improving positioning accuracy. With the rapid advancement of UAV technology, the range of applications for UAVs continues to expand. However, the environmental and meteorological variations caused by altitude changes significantly impact UAV performance. In addition, UAVs are often required to perform tasks in diverse scenarios, further increasing the complexity and challenges of these missions. Therefore, evaluating the generalization ability of the proposed method across different flight altitudes and scenarios is important, as this directly impacts its feasibility and stability in practical applications. In this subsection, the generalization is evaluated.

\begin{figure}[!t]
\centering
\includegraphics[width=0.50\textwidth]{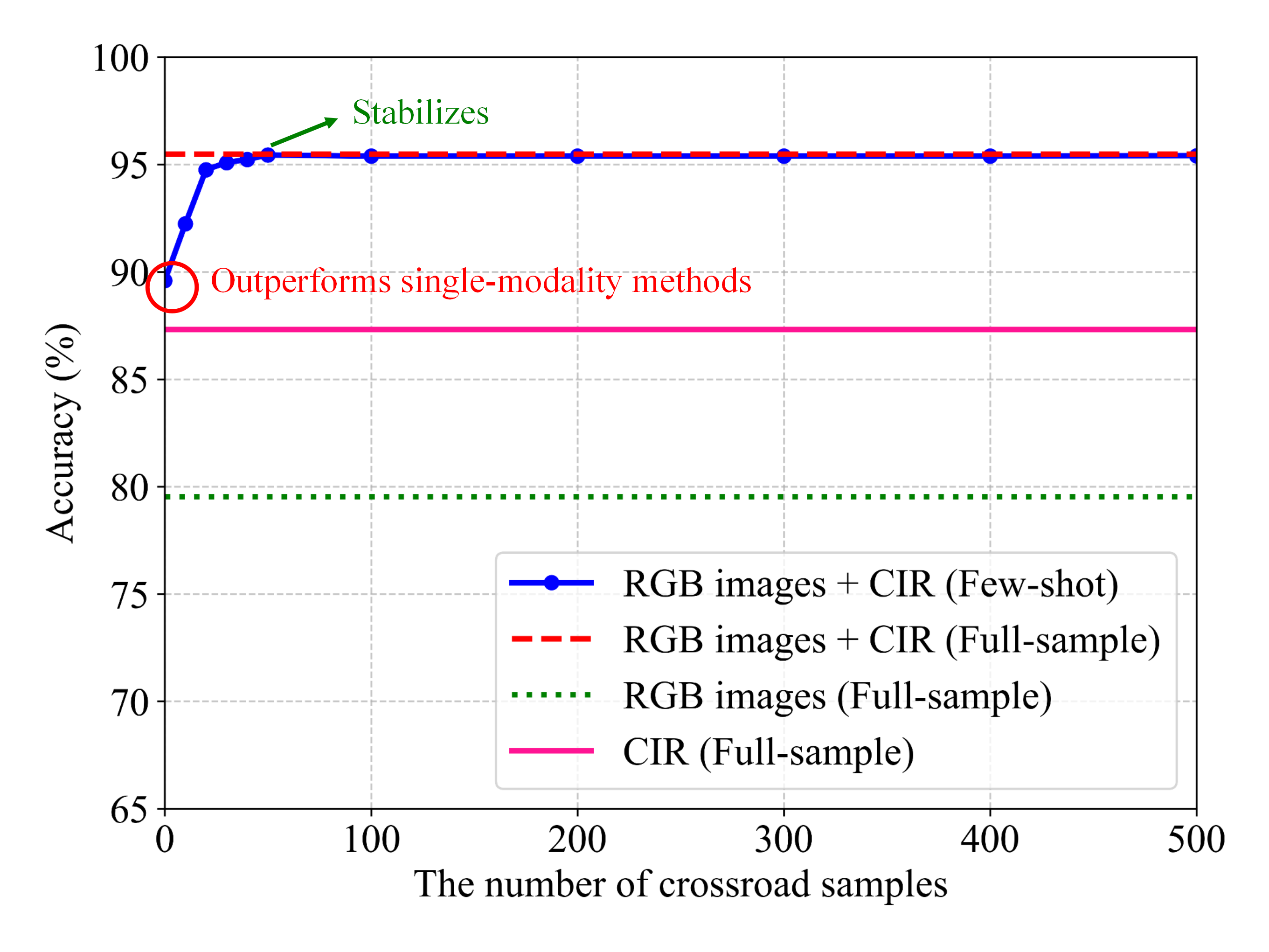}
\caption{Cross-scenario few-shot generalization performance of LoS/NLoS identification from wide lane to crossroad.}
\label{cross-scenario}
\end{figure}

To evaluate the cross-scenario and few-shot generalization capabilities of the model to adapt to diverse real-world conditions \cite{few-shot}, the model is first trained in one scenario and then fine-tuned utilizing a small number of samples in another scenario. Fig. \ref{cross-scenario} illustrates the few-shot generalization performance when the model is trained on wide lane scenario and evaluated on the crossroad scenario. Under the zero-shot training setup, the model trained on wide lane data achieves an accuracy of 89.59\% when tested in the crossroad scenario as shown by the blue line in Fig. \ref{cross-scenario}. The results indicate that the data from different scenarios share transferable features that are relevant to LoS/NLoS identification, which enables the model to achieve identification accuracies exceeding 89\% under the zero-shot training setup. Moreover, the complex distribution of buildings in the wide lane scenario enables the model trained on wide lane data to achieve high identification accuracy under zero-shot conditions.

As shown by the blue line in Fig. \ref{cross-scenario}, the LoS/NLoS identification accuracies between the wide lane and crossroad scenarios improve consistently as the number of target scenario samples utilized for fine-tuning increases. When the number of target samples for fine-tuning increases to 50, the accuracy of wide lane-to-crossroad stabilizes and approaches the full-sample training accuracy. Notably, in the wide lane-to-crossroad simulation, due to the greater complexity of the wide lane scenario, the fusion method surpasses the single-modality methods with zero target samples. The experimental results indicate that the proposed model can rapidly adapt to target scenarios and exceeds single-modality methods with limited sample data. The improvement is attributed to the complementary sensing and communication information, which reduces overfitting of single modal data to training scenarios.

\begin{figure}[!t]
	\centering
	\subfigure[]{\includegraphics[width=0.50\textwidth]{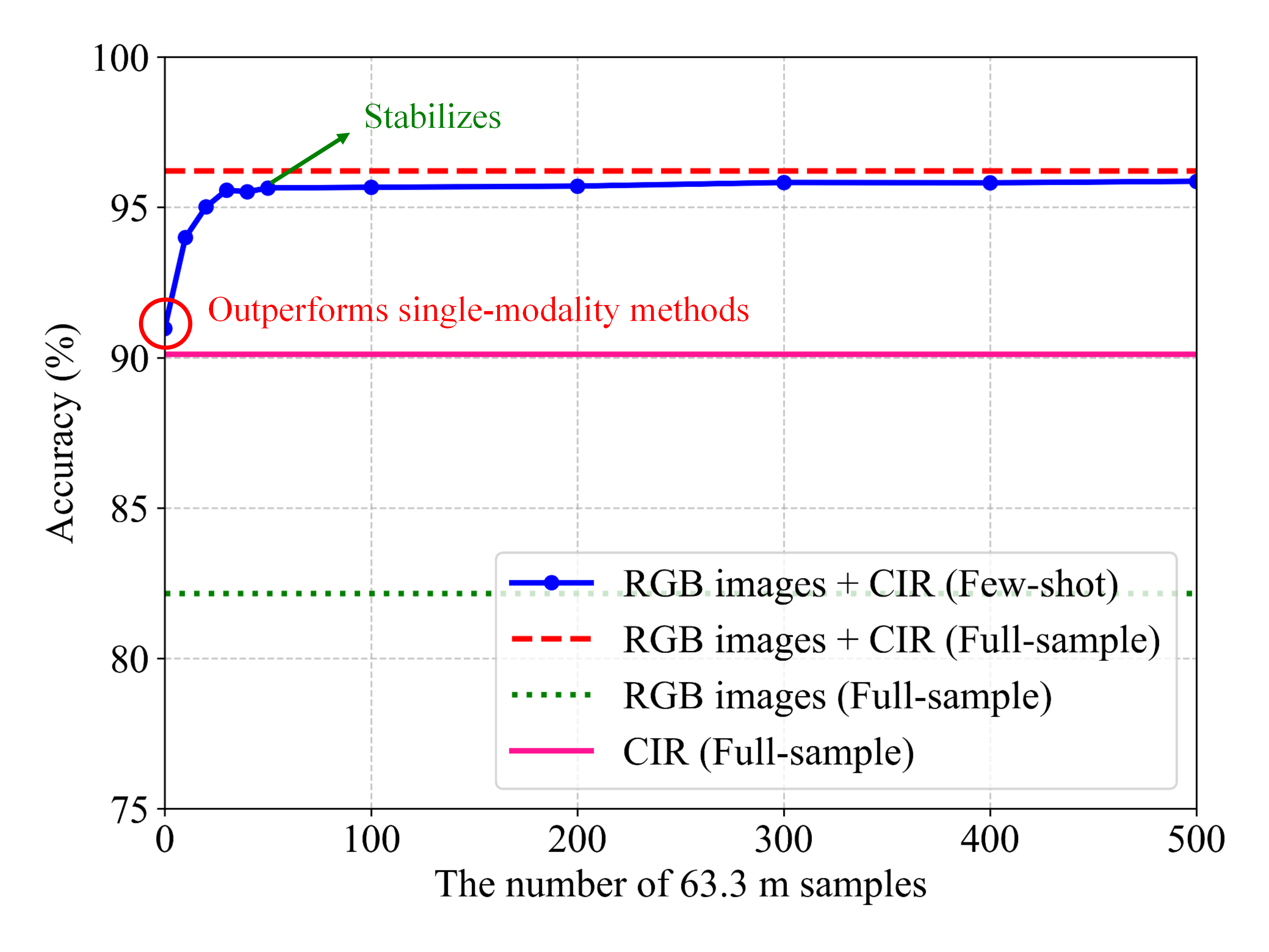}}
	\subfigure[]{\includegraphics[width=0.50\textwidth]{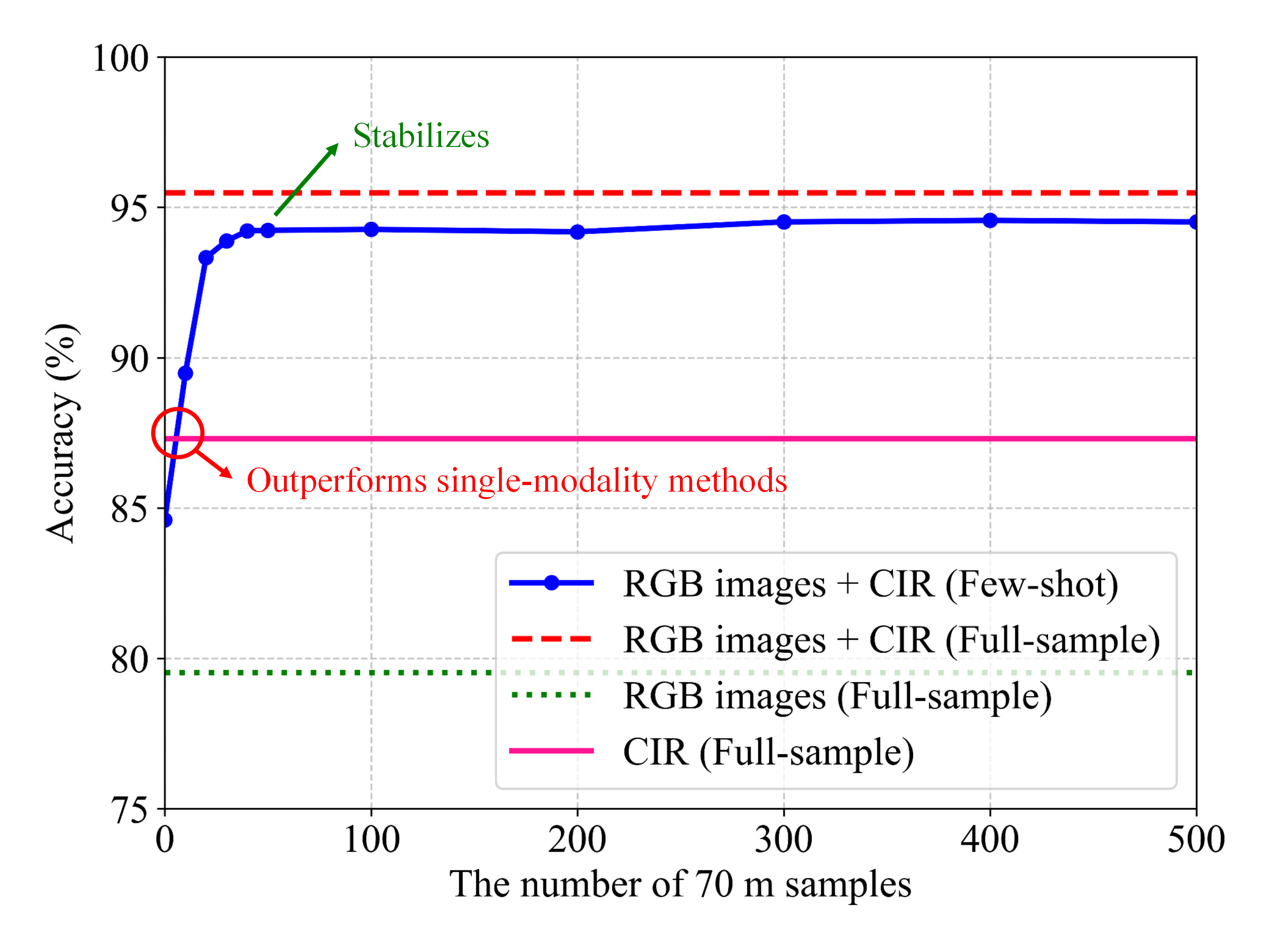}} 
	\caption{Cross-altitude few-shot generalization performance of LoS/NLoS identification in crossroad scenario. (a) 50 m to 63.3 m. (b) 50 m to 70 m.}
 \label{cross-altitude-50m}
\end{figure}

\begin{figure}[!t]
	\centering
	\subfigure[]{\includegraphics[width=0.50\textwidth]{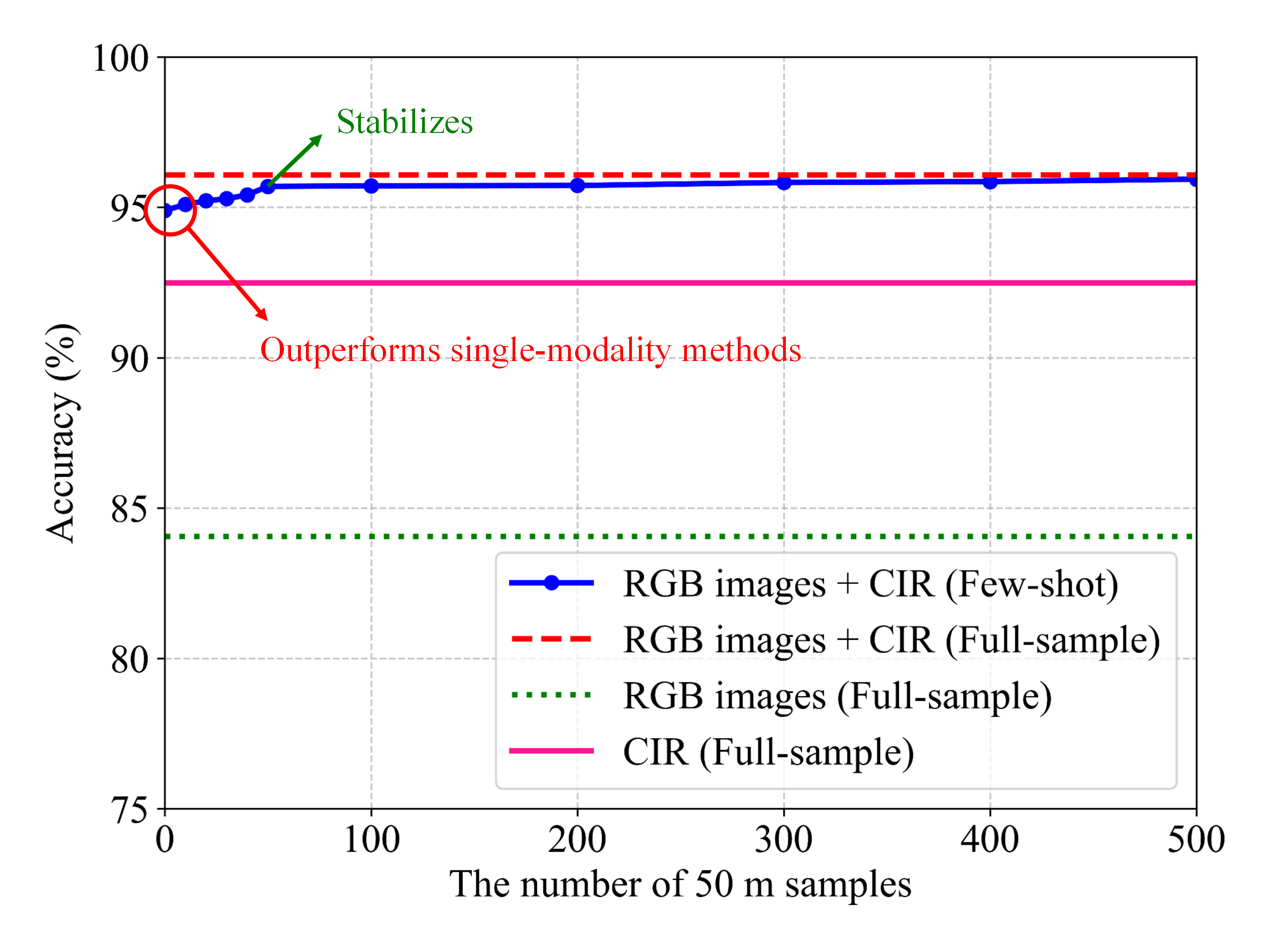}}
	\subfigure[]{\includegraphics[width=0.50\textwidth]{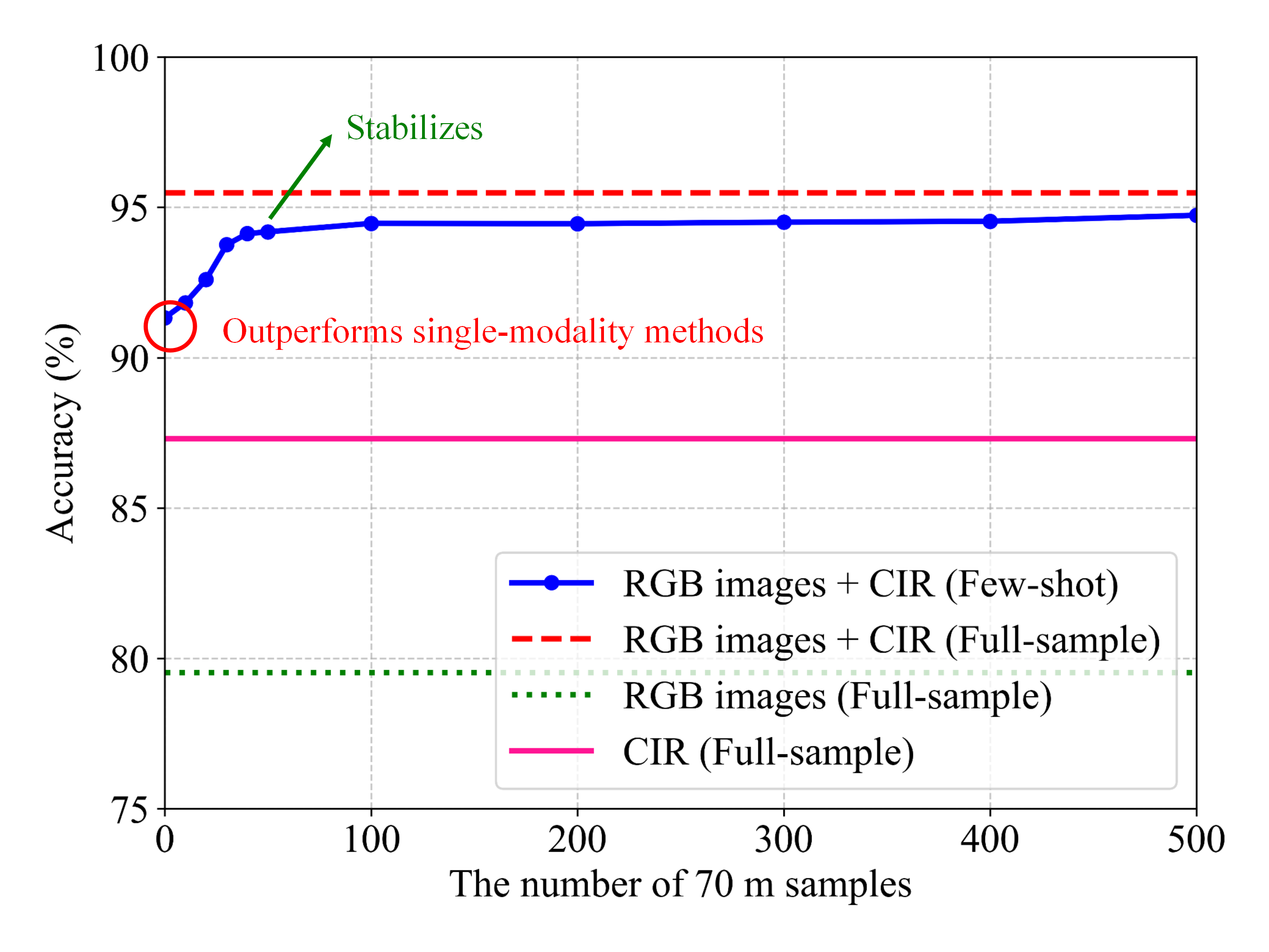}} 
	\caption{Cross-altitude few-shot generalization performance of LoS/NLoS identification in crossroad scenario. (a) 63.3 m to 50 m. (b) 63.3 m to 70 m.}
 \label{cross-altitude-63.3m}
\end{figure}

\begin{figure}[!t]
	\centering
	\subfigure[]{\includegraphics[width=0.50\textwidth]{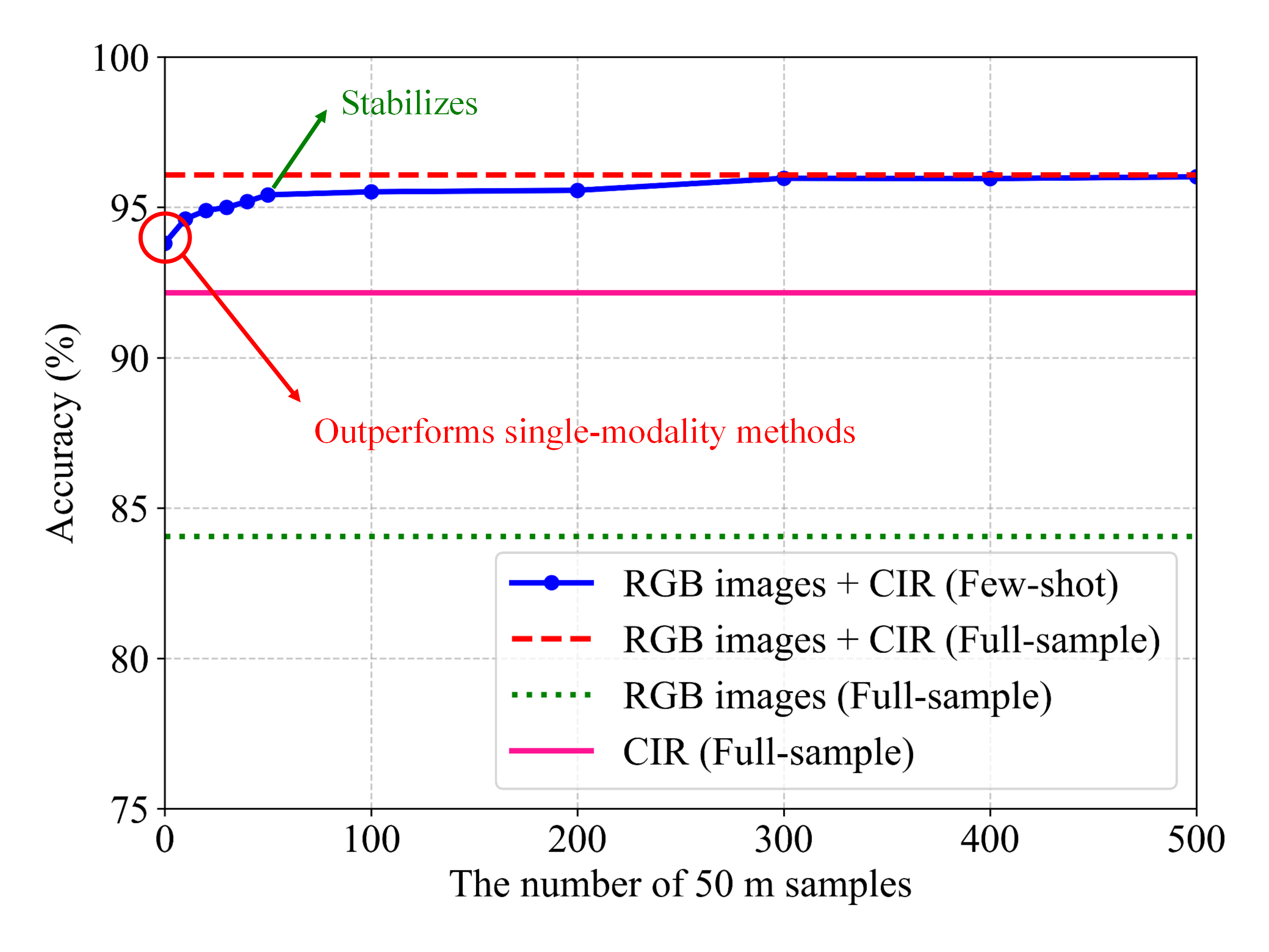}}
	\subfigure[]{\includegraphics[width=0.50\textwidth]{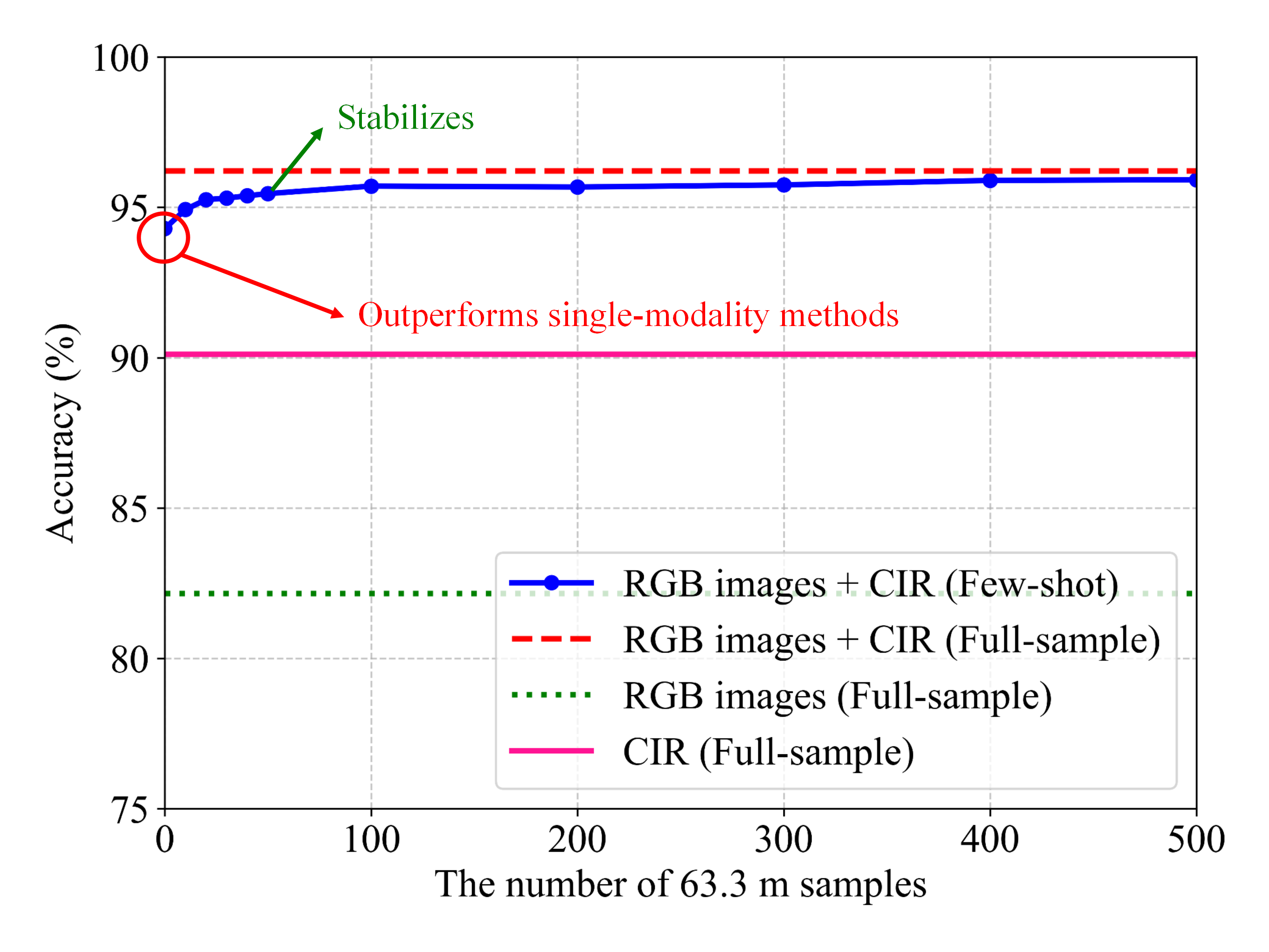}} 
	\caption{Cross-altitude few-shot generalization performance of LoS/NLoS identification in crossroad scenario. (a) 70 m to 50 m. (b) 70 m to 63.3 m.}
 \label{cross-altitude-70m}
\end{figure}

To evaluate the cross-altitude few-shot generalization capabilities, the model is trained utilizing data from a single flight altitude and subsequently fine-tuned and tested at other altitudes. Fig.~\ref{cross-altitude-50m}, \ref{cross-altitude-63.3m}, and \ref{cross-altitude-70m} present the cross-altitude few-shot generalization results obtained in the crossroad scenario.

As shown in Fig.~\ref{cross-altitude-50m}, under zero-shot training conditions, the model trained with 50 m data achieved identification accuracies of 90.98\% (blue line in Fig.~\ref{cross-altitude-50m}(a)) and 84.61\% (blue line in Fig.~\ref{cross-altitude-50m}(b)) when tested at 63.3 m and 70 m, respectively. All accuracies exceed 84\%, indicating that the model learned transferable and height-invariant LoS/NLoS features. At the test altitude of 63.3 m, the model achieves higher identification accuracy than the single-modality methods, even when no target samples are provided. At 70 m, the model surpasses the single-modality methods with fewer than ten target samples. When tested at 70 m, the identification accuracy shows a significant decline compared to that at 63.3 m, indicating that the increase in altitude difference leads to reduced identification performance. The result can be attributed to the fact that larger altitude variations cause significant differences in the propagation environment, thereby making cross-altitude generalization more difficult. After the introduction of a small number of target-altitude samples, the identification accuracy rises rapidly. When the number of target samples reaches approximately 50, the accuracy becomes stable and approaches the level achieved through full-sample training. These results suggest that the proposed method maintains robust cross-altitude performance under limited target-sample conditions, which is further validated by the following experiments.

As shown in Fig.~\ref{cross-altitude-63.3m}, when the model is trained utilizing data collected at 63.3 m, it achieves 94.20\% accuracy (blue line in Fig.~\ref{cross-altitude-63.3m}(a)) when tested at 50 m and 91.33\% accuracy (blue line in Fig.~\ref{cross-altitude-63.3m}(b)) when tested at 70 m. In all experimental settings, the identification accuracy remains above 90\%, indicating robust cross-altitude generalization capability. The accuracy achieves at the lower test altitude of 50 m is higher than that at 70 m. This can be attributed to the fact that data collected at higher altitudes capture more complex and diverse environmental features, which contribute to improved feature learning and enhance generalization performance when transferred to lower altitudes. Moreover, even in the absence of any target-altitude samples, the proposed fusion method outperforms the single-modality methods, demonstrating a strong zero-shot cross-altitude generalization ability. As target-altitude samples are progressively introduced for fine-tuning, the identification accuracy exhibits a steady upward trend. When the number of target-altitude samples reaches approximately 50, the accuracy converges and closely approaches the performance achieved under the full-sample training setting.

As illustrated in Fig.~\ref{cross-altitude-70m}, the model trained utilizing 70 m data achieves 93.61\% (blue line in Fig.~\ref{cross-altitude-70m}(a)) and 94.60\% (blue line in Fig.~\ref{cross-altitude-70m}(b)) accuracy when tested at 50 m and 63.3 m, respectively. In both testing cases, the proposed fusion method consistently outperforms the single-modality methods, further demonstrating its robustness under cross-altitude conditions. These results once again confirm the strong cross-altitude transferability of the learned features. When the number of target-altitude samples for fine-tuning reaches approximately 50, the performance converges and closely approaches that obtained under full-sample training. Combined with the results obtained at other training altitudes, these simulation results demonstrate that the proposed fusion method possesses a strong few-shot generalization capability across varying flight altitudes, enabling reliable LoS/NLoS identification even under limited data conditions.

\subsection{Noise Perturbation Generalization}

\begin{figure}[!t]
	\centering
	\subfigure[]{\includegraphics[width=0.23\textwidth]{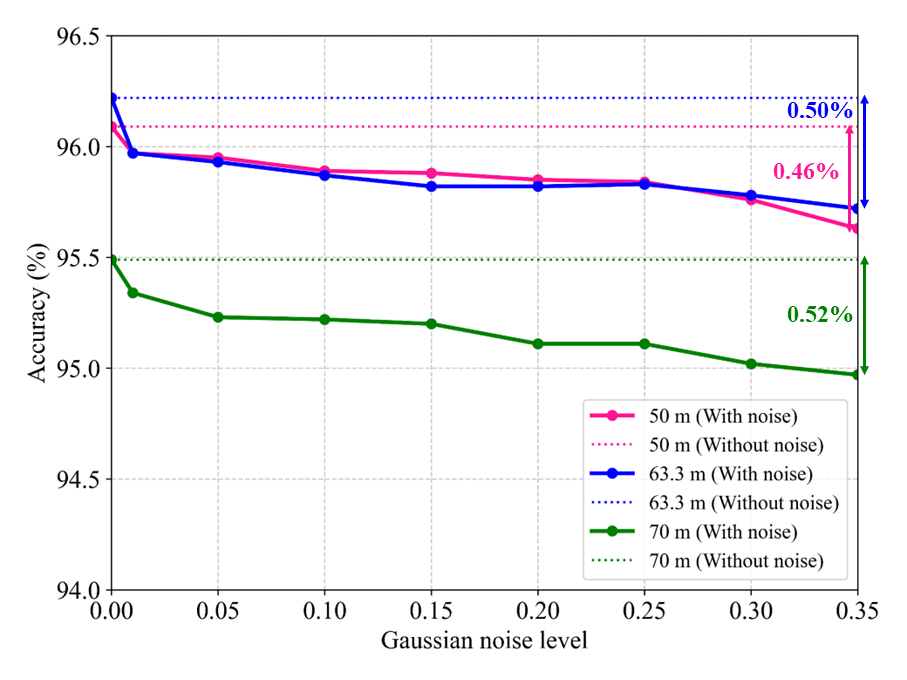}}
	\subfigure[]{\includegraphics[width=0.23\textwidth]{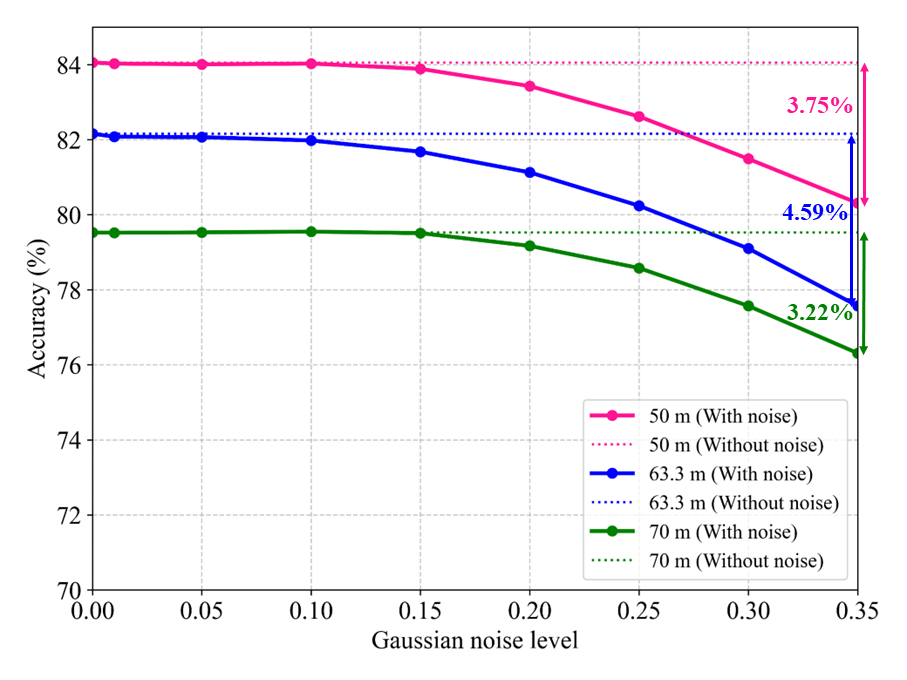}} 
	\caption{Generalization performance of LoS/NLoS identification under Gaussian noise added to RGB images at different altitudes in crossroad scenario. (a) The proposed fusion method. (b) The RGB-only method.}
 \label{noise}
\end{figure}

In practical data collection, UAVs typically operate at varying flight altitudes, which inevitably exposes them to various environmental conditions, such as weather changes and atmospheric interference. These factors introduce noise into captured images, which may affect identification accuracy. Therefore, in addition to strong cross-scenario and cross-altitude generalization, the method must demonstrate robustness against noise interference to ensure reliable performance in real-world applications \cite{noise}. Fig. \ref{noise} illustrates the performance of the trained model tested with data from three different flight altitudes under varying levels of Gaussian noise \cite{Gaussian}. As noise intensity increases, the identification accuracy gradually decreases. However, within a certain range of noise intensities, the drop in accuracy of the proposed method remains relatively small, demonstrating the robustness of the proposed method. Specifically, when Gaussian noise with a variance of 0.35 is added, the accuracy drops by only around 0.5\% compared to noise-free data. In contrast, under the same noise level, the RGB-only method suffers an accuracy drop of more than 3\%. The result indicates that the proposed method is capable of effectively resisting moderate noise interference. By learning global features from the images and fusing data from two modalities, the proposed method enhances its tolerance to noise, ensuring robust performance even in dynamic and unstable environments. Overall, the noise perturbation generalization results demonstrate the feasibility and reliability of the proposed method for practical UAV applications.

\subsection{Performance Evaluation of the Downstream Positioning Task}

%NLoS引入误差
\begin{figure*}[h]
\centering
\includegraphics[scale = 0.46]{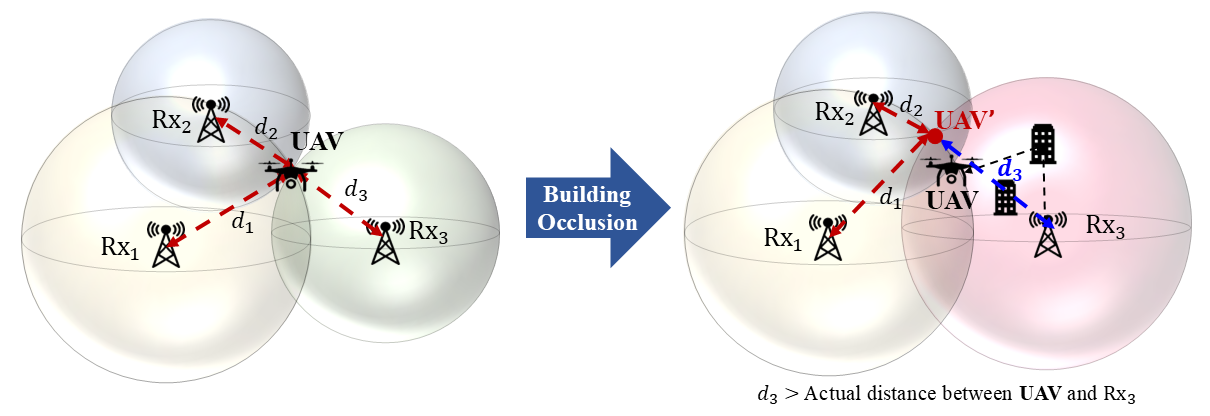}
\caption{ToA-based trilateration positioning and NLoS-induced error.}
\label{ToApositioning}
\end{figure*}

During UAV flight, the UAV continuously transmits both communication signals and captured RGB images to the ground Rxs in real time. The ground system then performs positioning based on the received data to assist the UAV in executing its tasks. Since the ToA of the signal can be obtained during data simulation, a ToA-based trilateration method, which has been widely used for UAV positioning in urban scenario, is utilized to perform UAV positioning \cite{UAVPOSITIONING}. As shown in Fig. \ref{ToApositioning}, the distances between Rxs and the UAV are calculated to realize positioning. The distances are calculated by
\begin{equation}
d_i = c \times ToA_i  
\end{equation}
% $\mathrm{Rx}_i$
where $c$ denotes the speed of light. Assuming the position of the UAV is $(x,y,z)$, and the known coordinates of each $\mathrm{Rx}_i$ are $(x_i,y_i,z_i)$. The geometric distance between the UAV and $\mathrm{Rx}_i$ satisfies
\begin{equation}
d_i = \sqrt{(x-x_i)^2+(y-y_i)^2+(z-z_i)^2}.
\end{equation}
The position of the target can be computed by minimizing the sum of the squares of a nonlinear cost function, i.e., the least-squares algorithm. The cost function can be formed by 
\begin{equation}
F(x)=\sum^N_{i=1}\alpha_i^2f_i^2(x)
\end{equation}
where $\alpha_i$ can be chosen to reflect the reliability of the signal received at the measuring unit $i$, $N$ denotes the total number of selected Rxs, and $f_i(x)$ is given as
\begin{equation}
f_i(x)=d_i-\sqrt{(x-x_i)^2+(y-y_i)^2+(z-z_i)^2}.
\end{equation}
The position estimate is determined by minimizing the function $F(x)$.

However, in densely built urban environments, a large number of NLoS paths exist. The ToA obtained from NLoS paths corresponds to the propagation time after undergoing reflections, refractions, and other interactions, which is typically longer than that of LoS propagation. Consequently, the distance estimated utilizing $d_i = c \times ToA_i$ becomes larger than the actual distance between the Tx and Rx. The NLoS paths introduce positive bias into distance estimation, resulting in positioning errors. As shown in Fig. \ref{ToApositioning}, when the signal path associated with $\mathrm{Rx}_3$ becomes NLoS, reflection introduces additional delay, causing the estimated distance $d_3$ to be overestimated. During the subsequent positioning process, this inflated distance constraint pulls the solution away from the UAV, resulting in a biased position estimate. Consequently, the UAV, which should be located at its real position, is erroneously positioned at a fictitious location $\mathrm{UAV}'$ dominated by the NLoS-induced bias.

%CDF图
\begin{figure}[h]
\centering
\includegraphics[scale = 0.40]{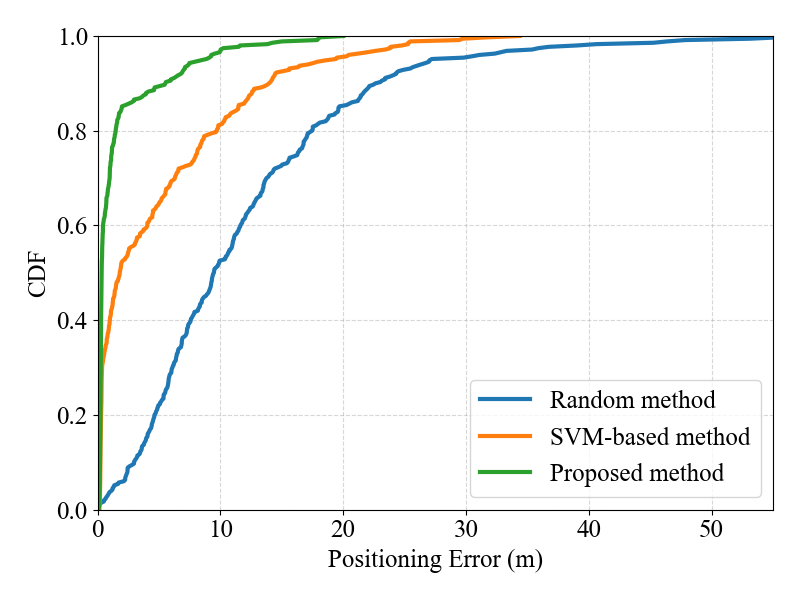}
\caption{CDF of positioning error based on different LoS/NLoS identification methods in crossroad scenario.}
\label{CDF}
\end{figure}

\begin{table}[!t]
\centering
\caption{Comparison of Identification Accuracy and Mean Positioning Error under Different LoS/NLoS Identification Methods}
\label{table4}
\renewcommand{\arraystretch}{1.3}
\setlength{\tabcolsep}{7pt}
\begin{small}
\begin{tabular}{l c c}
\toprule
\textbf{\makecell{LoS/NLoS \\Identification Methods}} & \textbf{\makecell{Identification \\Accuracy (\%)}} & \textbf{\makecell{Mean Positioning\\ Error (m)}}  \\ 
\midrule
Random Method \cite{trilateration} & - & 5.14  \\
SVM-based Method \cite{threeM}  & 89.90 & 2.36  \\
Proposed Method & \textbf{96.22} & \textbf{1.48} \\
\bottomrule
\end{tabular}
\end{small}
\end{table}

In typical urban scenarios, a large number of NLoS propagation paths arise due to building blockage and multi-path effects. In simulated crossroad scenario with a flight altitude of 63.3 m, a total of 1,078,185 LoS paths and 802,815 NLoS paths are generated. As a result, the positioning error caused by NLoS can be reduced through LoS/NLoS identification, and better identification performance leads to fewer errors. As shown in Table \ref{table4}, the overall positioning performance in the crossroad-63.3 m scenario is evaluated and compared under the three aforementioned LoS/NLoS identification methods. The random method selects three Rxs randomly for each snapshot to estimate the position, which results in an average positioning error of 5.14 m. For the SVM-based method, positioning is assisted by an SVM-based LoS/NLoS identification method with an identification accuracy of 89.90\%, which yields an average positioning error of 2.36 m. The proposed method identifies LoS/NLoS based on sensing and communication data and utilizes only LoS-identified Rxs for positioning, which results in an average error of 1.48 m. By effectively identifying and filtering out NLoS data, the proposed method has reduced the positioning error by approximately 70\% compared to the trilateration positioning error based on the random selection method. The results demonstrate that positioning methods assisted by LoS/NLoS identification methods significantly outperform the random method. Furthermore, owing to the higher identification accuracy, the proposed method yields better positioning precision than the SVM-based method.

To further evaluate the positioning performance, the cumulative distribution function (CDF) of positioning errors over the entire test set of the crossroad-63.3 m scenario is presented in Fig. \ref{CDF}. The horizontal axis represents the positioning error, while the vertical axis denotes the cumulative probability. It is observed that the proposed method consistently achieves lower positioning errors, as its CDF curve is located on the left side of the other two methods and increases more rapidly. This indicates that a larger proportion of positioning results falls within a small error range, demonstrating better stability and robustness. The proposed method reaches a cumulative probability of nearly 100\% at a smaller positioning error threshold compared to the other two methods. The CDF results further validate the effectiveness of the proposed LoS/NLoS identification method in improving UAV positioning performance.

\section{Conclusion}
In this paper, a new multi-modal sensing-communication integrated dataset in urban UAV-to-ground scenarios has been constructed, which includes RGB images, CIR matrices, and LoS/NLoS identification label matrices. Based on the constructed dataset, a novel sensing-assisted LoS/NLoS identification method for dynamic UAV positioning has been proposed. The proposed method has achieved identification accuracies of over 95\%. Compared with utilizing only RGB images or only CIR matrices, the identification accuracy of the proposed method has been improved by at least 3.59\%. Furthermore, the few-shot generalization capability of the proposed method across diverse scenarios and flight altitudes has been analyzed, as well as its robustness under noisy conditions. With exposure to fewer than 200 target samples from different scenarios or flight altitudes, the proposed method has achieved performance comparable to that of full-sample training and has demonstrated a strong few-shot generalization capability. Even under Gaussian noise with a variance of 0.35 applied to RGB images, the accuracy degradation remains approximately 0.5\%. The results have demonstrated that the model exhibits strong performance in practical applications involving varying scenarios, different UAV flight heights, and noise disturbances. By effectively identifying and filtering out NLoS data, the proposed method has reduced the positioning error by approximately 70\% in downstream tasks compared to the random method.

\end{document}